\documentclass{ws-ijqi}

\usepackage{color}

\begin{document}

\markboth{Aravinda, S. et al.}
{Orthogonal-state-based   cryptography  in  quantum mechanics and local post-quantum theories}

%
\catchline{}{}{}{}{}
%

\title{Orthogonal-state-based cryptography in quantum mechanics and 
local post-quantum theories}

\author{S. Aravinda}
\address{Poornaprajna Institute of Scientific Research, \\
4 Sadashivnagar Bangalore, India \\ aru@poornaprajna.org}
\author{Anindita Banerjee}
\address{Bose Institute,Centre of Astroparticle Physics and Space Science, EN Block, Sector - V, Salt Lake,
Kolkata, India\\ aninditabanerjee.physics@gmail.com}
\author{Anirban Pathak}
\address{Jaypee Institute of Information Technology,
A-10, Sector-62, Noida, India\\ anirban.pathak@jiit.ac.in}
\author{R. Srikanth}
\address{Poornaprajna Institute of Scientific Research, \\ 4 Sadashivnagar, 
Bangalore- 560080, India \\ srik@poornaprajna.org}
\maketitle


\begin{abstract}
We introduce  the concept of cryptographic  reduction, in analogy
with a  similar concept in  computational complexity theory.   In this
framework, class $A$ of crypto-protocols reduces to protocol class $B$
in a  scenario $X$,  if for  every instance  $a$ of  $A$, there  is an
instance $b$  of $B$ and  a secure transformation $X$  that reproduces
$a$ given $b$,  such that the security of $b$  guarantees the security
of  $a$.   Here  we  employ  this reductive  framework  to  study  the
relationship between  security in  quantum key distribution  (QKD) and
quantum secure  direct communication  (QSDC).  We show  that replacing
the streaming of independent qubits in  a QKD scheme by block encoding
and transmission (permuting the order  of particles block by block) of
qubits, we can construct a QSDC  scheme.  This forms the basis for the
\textit{block reduction} from a QSDC class of protocols to a QKD class
of protocols, whereby if the latter  is secure, then so is the former.
Conversely, given a secure QSDC protocol, we can of course construct a
secure QKD scheme by transmitting a  random key as the direct message.
Then the  QKD class of protocols  is secure, assuming the  security of
the QSDC  class which it  is built from.  We  refer to this  method of
deduction of security for this  class of QKD protocols, as \textit{key
  reduction}.    Finally,   we   propose   an   orthogonal-state-based
deterministic key distribution  (KD) protocol which is  secure in some
local  post-quantum  theories.   Its   security  arises  neither  from
geographic splitting of a code  state nor from Heisenberg uncertainty,
but from post-measurement disturbance.
\end{abstract}

\section{Introduction}

A protocol  of secure quantum  key distribution (QKD) was  proposed in
1984 by Bennett and Brassard \cite{bb84}. Since then several other QKD
protocols               have               been               proposed
\cite{ekert,b92,GisinsReview,vaidman-goldenberg,guo-shi,N09,ZWT0},  and the
notion of security has been considerably refined and strengthened.  It
is  now  established that  QKD  is  unconditionally secure,  while  by
contrast  any classical  cryptographic protocol  is secure  only under
some assumptions  about the hardness of  performing some computations.
This  important  feature  (unconditional  security) of  QKD  drew  the
attention of  the cryptography community.  However,  all the initially
proposed     protocols     of     secure     quantum     communication
\cite{bb84,ekert,b92,GisinsReview}  were  based  on  conjugate  coding
(i.e.,  encoding bits  using non-orthogonal  quantum states),  and the
applicability of these  initial protocols was limited to  QKD.  It was
soon realized  that quantum resources  can be used to  implement other
cryptographic tasks and that it  is possible to construct protocols of
secure quantum communication using orthogonal states.

Specifically, on  the one hand, conjugate-coding-based  protocols were
proposed for quantum secure direct communication (QSDC)
\footnote{In a QKD protocol, a  key is distributed first using quantum
  resources  and  then the  key  is used  later  for  encryption of  a
  message, whereas no such intermediary key is required in QSDC, which
  uses  quantum resources  to enable  legitimate users  to communicate
  directly without  establishing a key.   Further, a QSDC  scheme does
  not  require  any  classical   communication  for  decoding  of  the
  information. There exists another class of schemes for direct secure
  quantum  communication  (i.e.,   quantum  communication  without  an
  intermediary generation  of key) which  require additional classical
  information for  decoding of  the encoded information.  Such schemes
  are  referred  to  as  deterministic  secure  quantum  communication
  (DSQC).    }   \cite{ping-pong,lm05,CL},   quantum   secret  sharing
\cite{Hillery},  quantum dialogue  \cite{ba-an,qd} etc.  while  on the
other  hand,  a  few  protocols  with  orthogonal-state-based  quantum
cryptography were  proposed \cite{vaidman-goldenberg,guo-shi,N09}.  As
in the case of conjugate-coding-based protocols, the initial protocols
of         orthogonal-state-based         quantum        communication
\cite{vaidman-goldenberg,guo-shi,N09,Akshata1,cfMoreEfficientN09} were
also limited to QKD.

To    be     precise,    in     1995,    Goldenberg     and    Vaidman
\cite{vaidman-goldenberg}   proposed    first   orthogonal-state-based
protocol for deterministic QKD (called GV).  In 1997, Koashi and Imoto
\cite{koashi-imoto}  generalized  the  GV   protocol  and  proposed  a
protocol similar to GV protocol, but which obviates the random sending
time, a strict requirement of  the original GV.  Subsequently, in 1999
Guo and Shi \cite{guo-shi} proposed an orthogonal-state-based protocol
of  QKD  which  was  based  on the  principle  of  quantum  mechanical
interaction-free measurement \cite{Bomb  testing}.  Recently, in 2009,
Guo    and   Shi's    idea   was    extended   to    a   sophisticated
orthogonal-state-based counterfactual QKD  protocol by Noh \cite{N09},
which is now  famously known as N09 or the  counterfactual protocol of
QKD.   Later on,  in 2010  Sun and  Wen have  proposed a  modified N09
\cite{cfMoreEfficientN09} which is more efficient than N09.

All these orthogonal-state-based protocols
\cite{vaidman-goldenberg,guo-shi,N09,Akshata1,cfMoreEfficientN09} 
that were  proposed between 1995  to 2010 were only  theoretical ideas
and  limited to  QKD.  The  interest on  these  protocols considerably
increased in  the recent past as several  experimental realizations of
orthogonal-state-based protocols of QKD  were reported between 2010 to
2012      \cite{Marco-ground     breaking,GV-experiment,N09-expt-1,N09
expt-2,N09  expt-3}.  Motivated  by  these developments,  in last  two
years, some of the present  authors extended the existing protocols of
orthogonal-state-based QKD  to obtain orthogonal-state-based protocols
for     quantum     key     agreement     (QKA)     \cite{qka-chitra},
QSDC   \cite{beyond-gv,preeti-arxiv},   DSQC   \cite{dsqc-ent   swap},
counterfactual                                              certificate
authentication \cite{cf-srikanth-cert-authentication}, etc.

In most  of our recent works on  orthogonal-state-based secure quantum
communication, we have used  multi-qubit states and block transmission
after  applying a  permutation  operator  $\Pi$ on  the  qubits to  be
transmitted.    Specifically,    $\Pi$   scrambles   the    order   of
particles. This  procedure of permutation of particle  (PoP) was first
introduced  by Deng  and  Long  in 2003  \cite{CORE-03}  to propose  a
protocol of  ``controlled order rearrangement  encryption'' (CORE).  A
detailed  description of  this technique  and  a short  review of  the
orthogonal-state-based protocols that use  this technique can be found
in Refs. \cite{our review paper,my  book}.  Our recent works using PoP
and multipartite orthogonal states suggest that any cryptographic task
that can be performed using  conjugate coding can also be performed by
solely using orthogonal states \cite{our review paper}.

This   observation   in  general   and   successful  construction   of
orthogonal-state-based QSDC  \cite{beyond-gv,preeti-arxiv} protocol in
particular lead to  a couple of questions: (i) How  is the security of
QSDC  protocols  connected to  that  of  QKD  protocols?  (ii)  Is  it
possible to design  orthogonal-state-based protocols of secure quantum
communication  in  local post-quantum  theories  (say, in  generalized
local  theory (GLT), generalized  probabilistic theory  (GPT) or  in a
generalized non-signaling theory (GNST)) \cite{Barrett07}? (iii) Is it
possible to  design orthogonal-state-based quantum  device independent
protocols of QSDC and QKD.

The present  paper aims  to answer the  first two questions.   In what
follows, we answer the first question and show that we can construct a
QSDC  scheme  by  replacing  qubit  streaming in  a  QKD  scheme  with
block-encoding of qubits.  This reduces the security proof of the QSDC
scheme to the  security proof of the QKD scheme, in  the sense that if
the QKD scheme is  secure, then so is the QSDC scheme  built on top of
it using PoP.   This reduction scheme is referred  to as \textit{block
reduction}.  Similarly, it  is shown that the security  proof of a QKD
scheme reduces to that of a QSDC  scheme, in the sense that if we have
a secure QSDC scheme, then we  can always distribute a random key in a
secure manner using that scheme.  This reduction procedure is referred
to as \textit{key reduction}.   Further, we answer the second question
by  providing  an  orthogonal-state-based deterministic  QKD  protocol
which is secure  in some post-quantum theories (namely  in GPT, GLT or
the   local  part   of   GNST)   and  note   that   security  of   the
orthogonal-state-based protocol valid  in post-quantum theories arises
not from uncertainty, but from post-measurement disturbance.

The remaining part  of the paper is organized  as follows.  In Section
\ref{sec:crypto},   we  introduce  the   idea  of  cryptographic
reduction, inspired by a  similar idea in computability and complexity
theory, whereby the  security of a class of  protocols is derived from
another related to it by  means of a secure transformation. In Section
\ref{sec:Block-reduction}  we  describe  one   of  the  two  kinds  of
reductions considered here: \textit{block reduction}, while in Section
\ref{sec:Key-reduction} we study a reducton in the opposite direction,
called        \textit{key        reduction}.        In        Sec.
\ref{sec:Deterministic-QKD-in},  we  show that  orthogonal-state-based
protocols   of  secure   communication  can   be  designed   in  local
post-quantum   theories,  too.    Specifically,  the   possibility  of
orthogonal-state-based secure communication in a GLT and in local part
of  a  GNST  is  established.    Finally  we  conclude  the  paper  in
Sec. \ref{sec:Conclusions}.

\section{Cryptographic Reduction \label{sec:crypto}}

The concept of reduction in computability theory and complexity theory
is a basic tool by  which the (efficient) (un)solvability of different
problems   can   be   comparatively   deduced.    For   example,   the
\textit{Turing  reduction} from  problem  $A$ to  problem  $B$ can  be
regarded as an  algorithm to solve $A$, assuming that  an algorithm to
solve $B$ is given. In complexity theory, \textit{Cook reduction} from
problem $A$  to $B$  is a  polynomial-time algorithm  that efficiently
solves instances of  problem $A$ given an oracle that  solves $B$ in a
single time-step.

A comparison of QSDC  and QKD suggests that it is  useful to define an
analogous concept  of reduction  for cryptography.   The cryptographic
reduction in scenario $X$ from  crypto-protocols of class $A$ to those
of class $B$ is  a secure procedure $X$ by which  given a protocol $a$
in class $A$, there is an instance $b$ of class $B$, such that $a$ has
been  created from  $b$ by  using $X$,  and $a$  is secure  if $b$  is
secure. We discuss two scenarios $X$ below. In one, $X$ represents the
task of  block encoding qubits,  whereby a  fixed number of  qubits is
taken as a  block, re-arranged according to a  random permutation, and
then transmitted  as a block; this  gives block reduction from  a QSDC
class to a QKD class of  protocols.  Another scenario is one where $X$
represents  the substitution  of  a  random bit  string  for a  direct
message;  this  gives  key  reduction, which  works  in  the  opposite
direction to block  reduction.  These two reductions  are discussed in
the following two subsections.

\subsection{Block reduction\label{sec:Block-reduction}}

In  classical cryptography, block ciphers and  stream ciphers are
two  frequently used  algorithms for  encryption. In  the  former, one
employs  a symmetric  key  to  transform a  block  of fixed-length  of
bits.  For example,  the  key may  scramble  the bits  according to  a
certain permutation. By contrast,  a stream cipher combines individual
bits or  bytes of a  plaintext message with  a random key  stream.  By
default, QKD employs the  sequential streaming of individual qubits or
entangled states,  while QSDC conventionally  employs the manipulation
of  blocks   of  qubits  by  re-arranging  the   particles,  and  then
transmitting those blocks of qubits en bloc.

Recently  it  was shown  how  the use  of  block  transmission and  an
order-rearrangement  technique  can  make  an  orthogonal-state  based
deterministic     two-qubit      QKD     protocol     suitable     for
QSDC  \cite{preeti-arxiv}.   The QSDC  protocol  presented there:  (a)
Alice prepares $3N$ singlet states.  Of these she applies a PoP scheme
$\Pi$  on  $N$ pairs  together  with  $2N$  singlet halves.   (b)  She
transmits   these  re-arranged   $4N$   particles  to   Bob  over   an
authenticated   quantum   communication   channel.   (c)   After   Bob
acknowledges their receipt, she  reveals the information to unscramble
the $N$ full pairs transmitted to  Bob, who measures these in the Bell
basis  to determine the  error rate  $e$ by  a public  discussion with
Alice.  If  the error rate is  too high, they abort  the protocol run.
(d)  Alice encodes  a $2N  h(e)$-bit  message into  a classical  error
correcting  $[2N, 2N  h(e)]$-code, and  encodes this  onto $N$  of the
remaining  $2N$  qubits by  quantum  dense-coding.   Here  $h$ is  the
Shannon  binary entropy.  Alice  transmits these  $2N$ qubits  to Bob.
(e)  After Bob's  acknowledgment,  Alice identifies  the check  pairs,
which are then used to confirm  that the error rate remains not larger
than $e$. Upon confirmation,  Alice reveals the information that would
allow  Bob to  pair the  remaining  $N$ particles  with their  partner
particles, in order that the message may be decoded.

A QKD  scheme is  secure if  there is  a real  number $e_0$  such that
$0 \le e_0 \le 1$ and the observed error rate $e$ satisfies
\begin{equation}     
e\le e_{0}~\Rightarrow~I(A:B)\ge I(A:E).
\label{eq:secu}
\end{equation} 
This ensures that if Eve is restricted only to attack the (memoryless)
communication channel  (and the  devices and  initial states,  both of
which are  assumed to be well  characterized), then Alice and  Bob can
extract  some   secret  bits   via  key  reconciliation   and  privacy
amplification \cite{CK78}.

The   idea   behind   Eq.     (\ref{eq:secu})   is   essentially   the
information-vs-disturbance  trade-off in  quantum information  theory.
In a  QKD protocol, Alice generates  a random classical bit  or dit (a
$d$-level number) $s_j$.  She encodes the bit (or dit) in an entangled
state $|\phi^{(j)}\rangle_{S}$ and \textit{streams} (i.e, sequentially
transmits) $N$ such entangled states to Bob.  Eve prepares probe $P_j$
in the initial state $|\psi\rangle_{P}$.  When Alice's system $S_j$ is
transmitted  towards Bob,  Eve  executes the  interaction $U$  between
$P_j$ and $S_j$, producing the entangled state
\begin{equation}
\rho_{P}^{(j)}=
\textrm{Tr}_{S} \left[U^{\otimes N}\left( |\Phi\rangle_{S} |\Psi\rangle_{P}
\langle\Phi|_{S} \langle\Psi|_{P}\right)U^{\otimes N\dag}\right],
\label{eq:probe}
\end{equation}
where    $|\Phi\rangle_S    =   \bigotimes_j|\phi^{(j)}\rangle$    and
$|\Psi\rangle_P  =  |\psi\rangle^{\otimes  N}$. Based  the  subsequent
classical   communication  between   Alice  and   Bob,  Eve   measures
$\rho_{P}^{(j)}$  in a  suitable  basis to  extract information  about
$s_j$.   Here $U$  is optimal  in the  sense that  it maximizes  Eve's
information about  classical secret $s_j$  for a given  observed error
rate $e$.

In  the PoP-version  ($\mathcal{P}^\Pi$)  of  a cryptography  protocol
$\mathcal{P}$, because the streaming  is replaced with block-coding of
qubits,  Eve  does  not  know during  their  transit  which  entangled
particles are  partnered with  which.  She  replaces probes  $P_j$ and
with  a  single  master  probe   $P^\prime$,  and  replaces  $U$  with
$U^\prime$, a unitary  that interacts all $N$  qubits with $P^\prime$.
Here    the    primes    indicate    corresponding    quantities    in
$\mathcal{P}^\prime$.  Since  the PoP-version is  the same as  the old
protocol  with particle  re-arrangement,  therefore to  gain the  same
amount  of  information,   Eve  must  generate  a   greater  level  of
system-probe  entanglement  to  accommodate  the  various  permutation
possibilities, and correspondingly effect  greater channel noise.  For
any  fixed $I(A{:}E)  =  I^\prime(A{:}E)$, the  error rate  $e^\prime$
generated in  the PoP version  will be greater than  $e$.  Conversely,
for a  fixed error  $e =  e^\prime$, we  must have  $I^\prime(A{:}E) <
I(A{:}E)$.  

On the  one hand, this  means that the error  threshold ($e^\prime_0$)
till which $\mathcal{P}^\Pi$ remains secure  as a QKD protocol will be
larger, i.e.,  $e_0^\prime > e_0$.  The  quantitative determination of
$e_0^\prime$ will be  an interesting, if difficult,  question.  On the
other hand, if $e = e^\prime  = e_0$, then $I^\prime(A{:}E) < I(A{:}E)
= I(A{:}B) = I^\prime(A{:}B)$.  In fact, since there are exponentially
many ways to permute $N$ particles, $I^\prime(A{:}E)$ vanishes rapidly
as $N$  increases \cite{preeti-arxiv,CORE-03}.  This can  be expressed
as the QSDC condition
\begin{equation}    e\le
e_{0}~\Rightarrow~\lim_{N\rightarrow\infty}I(A:B)>I^\prime(A:E)\approx0,
\label{eq:secu0}
\end{equation}
which may be contrasted with the QKD condition (\ref{eq:secu}).  Since
$I^{\prime}(A:E)$ asymptotically  vanishes, Alice can  directly encode
her message, allowing  the sender to use a QSDC  protocol instead of a
QKD protocol.

Our   result    implies   that    given   a    QSDC   protocol
($\mathcal{P}^{\Pi}$)  obtained via  replacing stream  transmission by
block  transmission   in  a   QKD  protocol  $\mathcal{P}$,   then  if
$\mathcal{P}$ is secure, so is $\mathcal{P}^\Pi$ asymptotically at the
threshold  error rate  determined  by $\mathcal{P}$.   Define  by
$\mathfrak{S} \left( \mathcal{P}^{{\rm  QKD}}\right)$ the class of QKD
protocols  that transmit  data from  Alice to  Bob using  streaming of
individual qubits  or individual entangled states  of qubits.  Further
define  by $\mathfrak{S}  \left( \mathcal{P}^{{\rm  QSDC}}\right)$ the
class of QSDC protocols derived  therefrom using block coding. Thus to
every QSDC  protocol in this  class, we can associate  a corresponding
protocol  in  $\mathfrak{S}\left(\mathcal{P}^{{\rm  QKD}}\right)$,
whose  security guarantees that  of the  QSDC protocol.   We represent
this situation by:
\begin{equation}   \mathfrak{S}  \left(   \mathcal{P}^{{\rm
      QSDC}}\right)   \le_{B}   \mathfrak{S}  \left(   \mathcal{P}^{{\rm
      QKD}}\right).\label{eq:blockred}
\end{equation}              
This deduction  of the security  of a class  of QSDC protocols  from a
corresponding QKD class is block  reduction. A reduction that works in
the opposite direction is key reduction, which is described below.

\subsection{Key reduction \label{sec:Key-reduction}}

Given a secure QSDC scheme  $\mathcal{P}^{\Pi}$, it is obvious that it
can be  converted into a secure  QKD scheme, by transmitting  a random
key instead  of a  message.  Of  course, we must  assume that  the key
itself  is  truly  random  and  not vulnerable  to  attacks  like  the
known-plain-text attack,  etc.  Let the  class of  QKD protocols
obtained from  QSDC protocols in  this way--  by messaging a  key-- be
denoted    $\mathfrak{S}\left(\mathcal{P}^{{\rm   QKD}\prime}\right)$.
Thus  to  every  QKD  protocol  in this  class,  we  can  associate  a
corresponding  protocol  in   $\mathfrak{S}  \left(  \mathcal{P}^{{\rm
    QSDC}}\right)$,  whose   security  guarantees  that  of   the  QKD
protocol.  We represent  this situation, which works  in the direction
opposite to (\ref{eq:blockred}), by
\begin{equation}
\mathfrak{S} \left(\mathcal{P}^{{\rm QKD}\prime}\right) \le_{K} \mathfrak{S} 
\left(\mathcal{P}^{{\rm QSDC}}\right).
\label{eq:keyred}
\end{equation}
Broadly,  we  interpret  `$\mathcal{A}   \le_K  \mathcal{B}$'  as  the
key-reducibility   of   protocol    class   $\mathcal{A}$   to   class
$\mathcal{B}$,   meaning  that   we  obtain   a  secure   instance  of
$\mathcal{A}$  if   the  $\mathcal{B}$   counterpart  is   secure  and
furthermore a secure random number generator is available.

With a  slight modification, the above  argument can be extended  to a
quantum  key agreement  (QKA)  protocol \cite{qka-chitra}.   QKA is  a
quantum  key distribution  scheme in  which  Alice and  Bob must  both
contribute  equally   to  the  final  key.    Let  $\mathfrak{S}
\left(\mathcal{P}^{{\rm  QKA}\prime}\right)$ denote  the class  of QKA
protocols  derived from  QKD by  a  secure method  $X$.  For  example,
suppose that  the final reconciled  QKD key,  of even length  $m$, has
only Alice's  contribution. Then Bob publicly  announces $\frac{m}{2}$
coordinates, chosen randomly by him.  The final key is the secret bits
in these locations.   Clearly, if the QKD protocol is  secure, then so
is the derived QKA protocol.  We can express this situation by
\begin{equation}
\mathfrak{S} \left(\mathcal{P}^{{\rm QKA}\prime}\right) \le_{K} \mathfrak{S} 
\left(\mathcal{P}^{{\rm QSDC}}\right).
\label{eq:qkared}
\end{equation}
A QKA  protocol need not be derived  from a QKD protocol  in this way,
and  in that case,  such a  protocol is  not covered  in $\mathfrak{S}
\left(\mathcal{P}^{{\rm QKA}\prime}\right)$.

\section{Deterministic key distribution in 
local post-quantum theories \label{sec:Deterministic-QKD-in}}

GPT is  an operational  framework \cite{Barrett07}  that allows  us to
comparatively describe  a wide class of  theories, including classical
mechanics, quantum mechanics and post-quantum theories of the box-world
type  \cite{PR}.  Here  `operational'  means that  we  do not  concern
ourselves with  the state space  (such as Hilbert space),  but instead
only with  the vector space of  probabilities that can be  obtained by
performing  \textit{fiducial}  measurements  on allowed  states  in  a
theory.  A set of fiducial measurements $M = \{\mu = 0,1,\cdots,
\mu = K\}$ is  a  minimal and sufficient set whose outcome
statistics completely specify the state.

Therefore,  a state  in GPT  is  specified by  the probability  vector
obtained under  different fiducial measurements, $\mu =  0, 1, \cdots,
J$, with outcomes $\alpha=0,1,\cdots, K$:
\begin{equation}
\vec{{\bf P}}=\left(\begin{array}{c}
P(\alpha=0 | \mu=0)\\
P(\alpha=1 | \mu=0)\\
\vdots\\
\hline P(\alpha=0| \mu=1)\\
P( \alpha=1| \mu=1)\\
\vdots\\
\hline \vdots\end{array}\right),
\label{eq:gpt}
\end{equation}
where  $P(\alpha=k|\mu=j)$ is the  probability that  measuring $\mu=j$
yields      outcome      $\alpha=k$.       Normalization      requires
$\forall_j \sum_{\alpha=k}^{J}P(\alpha=k|\mu=j)=1$. 

In  classical  mechanics,  any  state can  be  specified  with  single
fiducial  measurement.  For  example,  the  state of  a  coin  can  be
represented by the toss probabilities, as
\begin{equation}
\vec{{\bf P}}=\left(\begin{array}{c}
P(\alpha=0|\mu=0)\\
P(\alpha=1|\mu=0)\end{array}\right),\label{eq:bit}
\end{equation}
where  $P(\alpha=0|\mu=0)$ is  the probability  of getting  $\alpha=0$
(i.e., head), and  $\mu=0$ is the measurement  implemented by tossing.
As  another  example, a  classical  particle  requires specifying  its
position and momentum,  but these two attributes can  be considered as
independent  systems  requiring  a single  fiducial  measurement.   By
contrast, a  (scalar) \textit{quantum} particle requires  two fiducial
measurements: position and momentum.

By measuring a qubit along $X$, $Y$, and $Z$ direction, the state of a
qubit  can  be  fully  represented, making  these  as  three  fiducial
measurements for the  state space of qubits in  quantum mechanics. For
example, the qubit with spin-up in  the $Z$ direction can be described
by
\begin{equation}
\vec{{\bf P}}=\left(\begin{array}{c}
P(0|X)\\
P(1|X)\\
\hline P(0|Y)\\
P(1|Y)\\
\hline P(0|Z)\\
P(1|Z)\end{array}\right)=\left(\begin{array}{c}
\frac{1}{2}\\
\frac{1}{2}\\
\hline \frac{1}{2}\\
\frac{1}{2}\\
\hline 1\\
0\end{array}\right),\label{eq:qbit}\end{equation}  where  $P(0|X)$  is
 the  probability  of  obtaining  spin  up by  measuring  in  the  $X$
 direction,   and  so   on.    

We note  that $\vec{P}$ in  Eq. (\ref{eq:qbit}) corresponds to  a pure
state in quantum mechanics. Now consider a local post-quantum theory--
a GLT-- with three fiducial  measurements but a larger state space, in
which  pure  states  $\vec{P}$   have  the  form  (\ref{eq:qbit})  and
$P(\alpha=k|\mu=j)$ is 0 or 1.   A pure qubit is represented a mixture
of such GLT pure states. A state in this GLT is called a
\textit{gbit} (for `generalized bit').

The gbit can quite generally be  of the type $J$-in-$K$-out, i.e., one
with $J$  fiducial measurements,  each with $K$  outcomes.  Obviously,
qubit is related most closely to a 3-in-2-out gbit.  Let us consider a
2-in-2-out   gbit,    whose   pure   states    are:   \begin{equation}
g_{0}=\left(\begin{array}{c} 1\\ 0\\
\hline 1\\
0\end{array}\right);~~g_{1}=\left(\begin{array}{c}
1\\
0\\
\hline 0\\
1\end{array}\right);~~g_{2}=\left(\begin{array}{c}
0\\
1\\
\hline 1\\
0\end{array}\right);~~g_{3}=\left(\begin{array}{c}
0\\
1\\
\hline 0\\
1\end{array}\right),\label{eq:gbit}\end{equation}   where  the   upper
(lower) pair refers to the state in fiducial property `$X$' (`$Z$').

An arbitrary gbit for our purpose is a convex combination of the above
four  elements.  For  example,  the $\pm1$  eigenstates  of the  qubit
observable     $X$    would     be     \textit{mixed}    states     of
gbits: \begin{equation}  |X+\rangle =  \left(\begin{array}{c} 1\\
0\\
\hline \frac{1}{2}\\
\frac{1}{2}\end{array}\right):=\frac{1}{2}(g_{0}+g_{1});~~~~
|X-\rangle := \left(\begin{array}{c}
0\\
1\\
\hline \frac{1}{2}\\
\frac{1}{2}\end{array}\right)=\frac{1}{2}(g_{2}+g_{3}).\end{equation}
A similar representation  follows if we also  include $Y$ measurements
in the picture, which we drop for simplicity.

Each of the above states may  be considered orthogonal in the sense of
pairwise  distinguishability,  i.e.,   there  is  a  measurement  that
deterministically  distinguishes  between  any  pair of  them.   E.g.,
$g_{0}$ and  $g_{1}$ are distinguished  by measuring $Z$,  while $g_0$
and  $g_2$  are  distinguished  by  measuring $X$.  These  gbits  lack
Heisenberg  uncertainty in that  both $X$  and $Z$  can simultaneously
take  definite values.   However, they  admit \textit{post-measurement
  disturbance}, whereby  performing one fiducial  measurement disturbs
the  statistics  of  the  other fiducial  measurements.  For  example,
measuring $X$ on  a qubit prepared in the  eigenstate of $Z$, disturbs
the   statistics  of   $Z$.   A   similar   unbiased  post-measurement
disturbance  is seen  in gbits.  For example,  measuring $X$  on $g_2$
deterministically returns  $\alpha=1$, but the  post-measurement state
will be an unbiased mixture  of $g_2$ and $g_3$.  Quantum measurements
have  both  uncertainty and  post-measurement  disturbance, while  GPT
states have only disturbance, and no uncertainty.

It  was  already known  \cite{Barrett07}  that  a  pair of  nonlocally
correlated gbits,  which form  a PR box\cite{PR},  can be used  for KD
using  a protocol  like the  Ekert protocol  \cite{ekert}, and  it was
conjectured that this  would be the case in  any non-classical theory.
The   method   we  earlier   used\cite{preeti-arxiv}   to  convert   a
deterministic QKD  protocol with Bell  states into a QSDC  protocol by
replacing  streaming   with  block-coding,  can  be   adapted  to  the
post-quantum  theories by replacing  Bell states  by PR  boxes.  Alice
prepares a  sequence of  PR boxes that  encode agreed-upon  bits.  She
permutes  the particles  use a  PoP configuration,  and  transmits the
resulting  particles to  Bob.  Bob  decodes them  after  receiving the
$\Pi$ information.

We  now show  that in  a  GLT, gbits  can be  used  as a  basis for  a
deterministic GLT-KD  (i.e., the GLT  version of QKD).  We  may regard
the  pure states  of  the  GLT `orthogonal'  if  there  is a  fiducial
measurement  that distinguishes  any pair  of them  in a  single shot.
This does not entail  that the set of all pure  states will be jointly
distinguishable. More generally, our result  applies to the local part
of   a  GNST   \cite{Barrett07}.    We  now   propose  the   following
deterministic  orthogonal-state-based GLT-KD  protocol (which  we call
GLT-2S):
\begin{description}
\item [{Encoding}] \textbf{and Sending:} Alice randomly and sequentially
generates  bit $x=0$  or  $x=1$.  In the  former  case, she  transmits
$g_{0}$ to Bob,  while in the latter case, $g_{3}$.  Note that $g_{0}$
encodes bit $x=0$ in both $X$ and $Z$, while $g_{3}$ encodes bit $x=1$
in both $X$ and $Z$.
\item [{Bob's}] \textbf{receipt:} Bob measures either $X$ or $Z$ in the
received gbit states, to extract the encoded bit deterministically.
\item [{Computing}] \textbf{error rate:}  Over a public channel, Alice
  and  Bob estimate  the error  rate on  the key  so extracted.   They
  publicly agree on certain coordinates  of gbits and observables ($X$
  or $Z$)  on those gbit  coordinates.  Bob announces the  outcomes on
  those coordinates.  If  too many of them are  mismatched, they abort
  the protocol.
\end{description}
Note that the coding in GLT-2S is not like conjugate coding in BB84 or
B92, because there is no uncertainty  between $X$ and $Z$ in GLT.  For
the same reason, while a public  announcement of bases would be needed
in  BB84,  here  none  is  required, and  furthermore,  the  raw  bits
generated are automatically  the sifted bits. Since  the coding states
are deterministically  distinguishable, and in that  sense orthogonal,
GLT-2S  may  be  considered  as  the  post-quantum  equivalent  of  the
Goldenberg-Vaidman  protocol  (GV) \cite{vaidman-goldenberg}.  On  the
other hand,  a single gbit  in GLT  is spatially localized,  making it
similar to BB84 in this sense, rather than to GV.

In  GLT-2S,   what  is   remarkable  is   that  security   comes  from
post-measurement disturbance, and not  from Heisenberg uncertainty. An
eavesdropper Eve (limited by the theory to only be able to perform the
stated  fiducial  measurements   \cite{DGV14})  can  deterministically
extract the encoded  bit by measuring either $X$ or  $Z$, but she will
disturb  the other  observable, which  can  be detected  in the  error
detection step.  If Eve measured $n$ gbits in either $X$ or $Z$ basis,
she  extracts $n$  bits of  information,  but she  disturbs the  other
fiducial  observable.  If  Bob measures  that during  the error  check
(which he  does with probability  $\frac{1}{2}$), he and  Alice detect
the attack with probability $\frac{1}{2}$.  Thus, the probability that
Eve   can   launch  this   attack   and   not   be  detected   is   $1
- \frac{1}{2}\cdot\frac{1}{2} = \frac{3}{4}$  per attacked gbit, which
drops   as  $\left(\frac{3}{4}\right)^{n}$   over  $n$   gbits.   This
exponential  drop  implies  unconditional   security  against  an  Eve
restricted to attacking single gbits.
In Eq. (\ref{eq:gbit}), if instead we use a GLT with 3 fiducial
measurements,  and a  similar GLT-KD  with two  states, this  would be
analogous   to  the   six-state  protocol   \cite{Bru98},   and  Eve's
corresponding probability to escape  detection will fall faster, given
by $p_{\rm esc} = \left(\frac{2}{3}\right)^{n}$. Generally, with a GLT
where gbits  are of  the type $J$-in-$K$-out,  $p_{\rm esc}  = \left(1
- \frac{J-1}{J}\frac{K-1}{K}\right)^n$, indicating  that in this sense
a protocol is more secure  in a theory with more fiducial measurements
and outcomes. 

\section{Conclusions and discussions\label{sec:Conclusions}}

Security  in   quantum  cryptography   has  contributions   from  both
Heisenberg  uncertainty  and  post-measurement disturbance,  while  in
GLT-2S, it  comes only from disturbance.   Intrinsic randomness, which
lies at the  heart of nonclassicality \cite{ASQIC},  can manifest both
in  uncertainty  and  post-measurement  disturbance.   Our  work  here
suggests  that  only  the  randomness concerned  with  disturbance  is
essential  to  cryptographic  security,  though  that  concerned  with
uncertainty can quantitatively modify the secure limit.

Inspired by  the analogous concept of  reduction in computability
theory and complexity  theory, we introduced in this  work the concept
of cryptographic reduction. In  particular, we defined block reduction
of QSDC class of protocols to QKD class of protocol, and key reduction
that works  vice versa.   These reductive methods  can be  used to
understand the relationship between the security of these two types of
crypto-tasks.  In condition  (\ref{eq:secu0}), since $I^\prime(A{:}E)$
asymptotically vanishes, we  expect that the condition  will hold good
even  when $e^\prime_0$  (the secure  threshold for  QSDC) drops  well
below the QKD threshold $e_0$,  and indeed, becomes arbitrarily small.
A  derivation  of  this  will  be  interesting  both  practically  and
foundationally.

As far as  we know, the protocol GLT-2S that  we proposed here appears
to  be   the  first  effort  to   design  an  (orthogonal-state-based)
cryptographic  protocol  for  a  \textit{local}  post-quantum  theory.
Related    work    has    been    either    about    a    post-quantum
non-signaling \cite{BHK05}  or signaling \cite{SASH0}  Eve attacking a
quantum   protocol,  or   a  protocol   in  a   post-quantum  nonlocal
theory  \cite{Paw10}.  GLT-2S   is  interesting  from  a  foundational
perspective  as it  provides  a  clearer insight  into  the origin  of
security in quantum mechanics, without reference to nonlocality.

\textbf{Acknowledgment:} 
SA acknowledges  support through the INSPIRE  fellowship [IF120025] by
the Department of Science and  Technology, Govt.  of India and Manipal
university  graduate programme.   AP and  RS thank  the Department  of
Science and Technology  (DST), India for support  provided through the
DST   projects   No.    SR/S2/LOP-0012/2010   and   SR/S2/LOP-02/2012,
respectively.

\end{document}